\documentstyle[prd,aps,preprint]{revtex} \tightenlines \newcommand
{\be}{\begin{eqnarray}} \newcommand {\ee}{\end{eqnarray}} %
\begin{document}

\draft
%\twocolumn[\hsize\textwidth\columnwidth\hsize\csname
%@twocolumnfalse\endcsname
\preprint{\vbox{
\hbox{UMD-PP-02-023}}}
\title{Mirror Matter as Self Interacting Dark Matter}
\author{ R. N. Mohapatra$^a$, S. Nussinov$^b$ and V. L. Teplitz$^c$}

\address{$^a$Department of Physics, University of Maryland, College Park,
MD-20742\\
$^b$ Department of Physics, Tel-Aviv University, Tel Aviv, Israel\\
$^c$ Department of Physics, Southern Methodist University, Dallas,
TX-75275 \\ and Office of Science and Technology Policy, Executive Office 
of the President, Washington D. C. -20502}

 \date{November, 2001}
\maketitle
\begin{abstract}
{It has been argued that the observed core density profile of galaxies
is inconsistent with having a dark matter particle that is collisionless
and alternative dark matter candidates which are self interacting may
explain observations better. One new class of self interacting dark
matter that has been proposed in the context mirror universe models of
particle physics is the mirror
hydrogen atom whose stability is guaranteed by the conservation of mirror
baryon number. We show 
that the effective transport cross section for mirror hydrogen atoms,
has the right order of magnitude for solving the ``cuspy'' halo problem. 
Furthermore, the suppression of
dissipation effects for mirror atoms due to higher mirror mass scale 
prevents the mirror halo matter from collapsing into a disk strengthening 
the argument for mirror matter as galactic dark matter.}
\end{abstract}

\section{Introduction}
It has recently been pointed out\cite{spergel} that the dark matter
particles constituting the galactic halo need to satisfy a new constraint
in order to avoid singular cusps\cite{singular}. One way to quantify this
constraint
is to demand that the mean free paths of these particles be less than
typical galactic sizes (say $0.1$ Mpc) i.e. $\lambda_{DM}\sim
\frac{1}{n_{DM} \sigma_{DM}}\leq 0.1 $ Mpc. This equation implies that
the typical cross section for the dark matter particle must be of order
$\sigma_{DM}\simeq \frac{m_{DM}}{GeV} 10^{-24}$ cm$^2$. This cross section
is large and it grows with the mass of the dark matter particle linearly.
Some favorite long standing candidates like the neutralino LSP of 50-100
GeV mass\cite{susy} would then need to have a scattering cross section of
$10^{-22}$ cm$^2$, a requirement which is not met by any of the exsiting
supersymmetric models. Barring some new long range interactions, such a
large cross section for particles of such high mass would be in conflict
with unitarity bounds\cite{chui} for pointlike dark matter particle
scattering via S-waves.

The growth of cross sections with mass is generic to solitonic structures
and it has been noted that $Q$-balls originally suggested in a different
context\cite{qball} can, for certain range of parameters, satisfy this
constraint\cite{kusenko}.

An alternative and natural candidate arises in mirror matter models where
it is postulated that there is a parallel standard model which duplicates
all the matter and forces and coexists, in our universe, with the familiar
standard model. The mirror and familiar particles in such models are only
connected by gravity\cite{lee,bere,volkas,sila}. In particular, the
asymmetric mirror model\cite{bere,teplitz} where both the weak scale as
well as the QCD scale in the mirror sector are about 20-30 times 
the corresponding scales in the the familiar sector has
been studied extensively in connection
with neutrino physics and explanation of the microlensing
events\cite{teplitz}\footnote{One can show that the asymmetry between
the two QCD scales owes its origin to the asymmetry between the weak
scales\cite{babu}. The main reason the first asymmetry follows from the
second is that the mirror quarks are much heavier than the familiar quarks
and therefore
decouple earlier from the evolution of the QCD couplings in the mirror 
sector. This helps to speed up the rise of the mirror QCD fine
structure constant.}. A particularly interesting feature of these models
is that the lightest mirror baryon $p'$ (in the form of mirror hygrogen
atom) is ideally suited to be the dark matter of the universe and as such
would be the dominant constituent of the dark halo of the galaxies. If 
the QCD scale parameter in the mirror sector $\Lambda'\approx 30 \Lambda$
the corresponding scale in the familiar sector, then $m_{p'}\approx 30
m_p$ and using the slightly lower reheat temperature of the mirror sector
(required to satisfy the BBN constraints arising from
$\gamma',\nu'_{e,\mu,\tau}$), we find that
$\frac{\Omega_{B'}}{\Omega_B}\approx
\left(\frac{T'}{T}\right)^3\frac{m_{p'}}{m_p}$. The BBN constraint
requires that $T'/T \approx (1/10.75)^{1/4}\sim 0.5$. Using this we get
$\Omega_{B'}\sim 4 \Omega_B$. For $\Omega_B\sim 0.05$ this would lead to
20\% dark matter and about 75\% dark energy. This is of the right order of
magnitude for the required fraction of the dark matter in the universe. 

A very important property that distinguishes mirror baryon from other dark
matter candidates is that mirror matter has self interaction. It was
suggested in a recent paper by two of the authors (R. N. M. and
V. L. T.)\cite{tep1}, that this might help resolve the core density
problem. We further pursue this question in the brief note.  
specifically taking account of the important distinction between total 
and transport cross sections, we show that the
parameters of the model suggested by considerations of $\Omega_{DM}$
yield scattering of mirror hydrogen atoms in the right range suggested in
Ref.1.

We then address the question of the shape of the dark halo if it
is made up of mirror dark matter particles. Mirror symmetry requires that
the coupling parameters in the mirror sector be identical to those of the
familiar sector. This has led to a suspicion that if halos were to be made
up of mirror baryons, they would collapse due to dissipation of their
transverse energy and become disk shaped, in contradiction to
observations. The point however is that even though the couplings are
identical due to mirror symmetry, the masses are different i.e. the mirror
matter masses are a factor of 30 or so higher. As a result, the processes
such as brehmsstrahlung responsible for dissipation of transverse energy
are down, preventing the collapse of mirror halo to a disk. All this makes
mirror baryons a viable dark matter candidate.

\section{Effective scattering crosssection for mirror hydrogen}
For small relative velocities of atoms of order $\beta_{virial}\sim
10^{-3}$, the total atom-atom elastic scattering crosssections are of
order $\pi R^2_{atom}$. For $H$ or $He$ atoms, $R_{atom}\sim 0.55$
angstroms leading to $\sigma_{HH}\simeq 10^{-16}$ cm$^2$. If we
take the mirror scale factor to be about $30-100$, then the Bohr radius of
the corresponding hydrogen atoms will scale inversely with it and will
give $\sigma_{H'H'}\simeq 10^{-19}- 10^{-20}$ cm$^2$. This value is
higher than the value apparently required for solving the core density
problem by a factor of 100-1000. The new observation in this note is that
 the naive use of the cross section is not adequate for our discussion and
there is indeed a substantial suppression factor which arises from a more
careful analysis.

The main point is that the cross section relevant for avoiding
the catastrophic accumulation of dark matter particle particles is {\it
not} the total elastic cross section, $\sigma_{el}$ but the transport
cross section, $\sigma_{tr}$,
to which large angle scattering contributes more strongly i.e.
\begin{eqnarray}
\sigma_{tr}= \frac{1}{4\pi}\int d\Omega
(1-cos\theta) \frac{d\sigma}{d\Omega}
\end{eqnarray}
For isotropic (say S-wave) or slightly backward hard sphere scattering,
$\sigma_{el}$ and $\sigma_{tr}$ are roughly the same. This is however not
the case for H-H or H'-H' scattering at relative velocity of $\beta
\approx 10^{-3}$. Here many partial waves upto $\ell_{eff} = m_Hv
r_{Bohr}\approx m_{H'} v r'_{Bohr}\approx 200$ contribute allowing for
strongly forward peaked elastic differential cross sections. To estimate
this cross section, note that: (i) large number of partial waves suggest a
quasi-classical WKB treatment; (ii) the collision virial velocity is
smaller than the velocity of the electron in the atom
i.e. $10^{-3}c\approx \beta_{virial} c < \alpha_{em} c$, where
$c\alpha_{em}$ is the velocity of the electron in the atom. Hence we can
adopt an adiabatic Born-Oppenheimer type approximation. The interatomic
potential can be computed for each atom-atom configuration denoted by
the impact parameter $b$ and the position of the $H'$ along its path
(assumed to be
a strightline) $z(t)$. The interatomic distance is then given by:
$R(t)= \sqrt{z^2(t) + b^2}$. The magnitude of the interatomic potential
$V_{HH}(R)\approx m_e \alpha^2_{em}\approx 27$ eV is about 20 times
smaller than the kinetic energy of the collision
$\frac{1}{2}m_H\beta^2\sim 500$ eV (the same ratio applies to the mirror
sector since both terms get scaled by a common factor). Hence for such
velocities, atoms are ``soft" and interpenetrate quite a bit. As we
indicate, the
scattering angle $\frac{\Delta p}{p}$ is approximately given by
$\frac{\Delta p}{p}\sim
\frac{V}{1/2m_e\beta^2}\simeq \frac{1}{20}$. The classical
deflection angle which may be appropriate here is
\begin{eqnarray}
\theta \approx \frac{\Delta p_y}{p}=\frac{\int F_y(z(t), b) dt}{p}\simeq 
2\int^{\infty}_0 \left[\frac{\partial V(\sqrt{z^2+b^2})}{\partial
y}\right]\frac{dz}{pv}
\approx \frac{2V}{mv^2}\sim \frac{V}{T}.
\end{eqnarray}
Hydrogen-Hydrogen scattering at KeV energies can be measured
experimentally and calculated with high accuracy. We believe that the
qualitative features of strong forward peaking and correspondingly reduced
transport cross section will still be manifest. Thus the
transport cross section, which is $\frac{1}{2}<\theta>^2\sigma_{el}$ will
be  about $10^{3}$ times smaller than the naive geometric
value. The transport
cross section for mirror hydrogen then is of order $10^{-22}$ cm$^2$,
which is close to the required value for self interacting dark matter.
These approximations are commensurate with the data and calculations on
cross section for
H-H scattering\cite{ornl}. For scattering to excited states, including
ionization, we would expect less forward peaking but smaller cross
section at KeV energies.

\section{Dissipation and shape of the mirror halo}
We next examine  the  dissipation time scale
which is important for understanding the shape of the mirror dark matter
halo. Mirror symmetry implies that mirror particles like ordinary ones are
dissipative, namely that energy can be lost by $\gamma'$ emission. For
the baryonic matter in galaxies, it is this process of energy loss
that causes the collapse to a galactic disc, which provides a lower energy
configuration with same total angular momentum. However if mirror matter
is to form a realistic, roughly spherical galactic halo, such disc
formation should not be
allowed. The time scale for the disc formation in our galaxy has been
estimated to be\cite{rees} to be equal the dynamical free fall time
$1/(G_N\rho)^{1/2}\approx 10^8$ years (using density of one proton per
cm$^3$)\footnote{This accident is crucial to the formation of the
disc. If the dissipation time was much larger, galaxy clusters would form
prior
to discs and if it were much shorter, the galaxy would likely fragment.}
The dominant dissipative process is thermal brehmstrahlung. Since the
latter scales like $m^{-2}$, for the mirror
baryons, the corresponding time scale would be
shorter by a factor $(m^2/{m'}^2)\sim 10^{-3}$. This slows the
relaxation time required to  form a disc to about 100 billion years, which
is way beyond the age of the universe.

Mirror star formation, as discussed in \cite{teplitz} does not depend on
brehmstrahlung, but rather on molecular cooling and is not affected by the
present discussion. We do however note, in that connection, that one can
derive from rather general principles an additional scaling rule (with
$\zeta \equiv m'/m$) for compact objects made of mirror matter, for the 
mass of the minimum cloud that will further fragment, as shown in the
appendix.

In conclusion, we have pointed out that, in mirror matter models, the
mirror hydrogen atom has all the right properties to be a self interacting
dark matter of the universe. In particular, we note that due to near
forward nature of the H-H scattering, the effective, relevant 
 transport cross sectionm is around
$10^{-22}$ cm$^2$, and is adequate to damp the core density of the dark
matter in galactic halos. We also point out that due to reduced
bremstrahlung cross section of mirror matter, dissipation processes get
weakened enough so that the dark matter does not become disc shaped.

\bigskip

We like to thank C. Canizares,
A. Kusenko and P. Steinhardt for discussion and  D. Schultz for valuable
information on the Hydrogen scattering. We are particularly indebted to
Jeremy Goodman for clarifying to us the issues
involved in fixing the time scale for galactic disc formation. The work of
R. N. M. is supported by the National Science Foundation grant number
PHY-0099544 and that of S. N. by the Israel Academy and partially by the
National Science Foundation.

\bigskip

\bigskip

\begin{center}
{\bf Appendix}
\end{center}

\bigskip

In this appendix, we discuss the scaling with $\zeta\equiv
\frac{m_{p'}}{m_p}=
\frac{m_{e'}}{m_e}$ for stellar masses. The first point to note that
for compact objects, the free fall time is much shorter than the
corresponding one for galaxies. Therefore, we can expect compact objects
to form in the mirror sector of the universe. Let us now
estimate the maximum mass of the
mirror cloud that will not further fragment. We follow the treatment in
Carroll and Ostlie\cite{carroll}. They ask that the cloud luminosity
needed to permit loss of potential energy $\Delta E \sim
\frac{3}{10}\frac{GM^2}{R}$ in the free fall time $t_{ff} \sim
(G\rho)^{-1/2}$, (to be called $L_{ff}\equiv \frac{\Delta E}{t_{ff}}$) be
less than the black body luminosity $L_{bb}= 4\pi R^2\sigma T^4 $.
This gives
\begin{eqnarray}
G^3 M^5 \approx T^8 R^9
\end{eqnarray}
where we have omitted all non-dimensional constants. We can combine it
with the equation that follows from the virial theorem i.e. $\frac{Gm_p
M}{R}
= kT$ to get
\begin{eqnarray}
M~=~ \left(\frac{T}{m}\right)^{1/4}\cdot \frac{1}{G^{3/2}m^2_p}
\end{eqnarray}
 This shows that the size of the
minimum fragmenting cloud scales like $\zeta^{-2}$(assuming that the
mirror temperature rises with $\zeta$ as is the case). In practice, we do
not expect the cooling processes to support fragmentation to such a low
mass, in view of cross section decrease with $\zeta$. Rather as pointed
out in \cite{teplitz}, we expect suppression of the processes that limit
accretion in familiar stars so that the mirror stars will tend to cluster
in mass near the maximum allowed stellar mass.

The point we believe should be noted is that the $\zeta^{-2}$ scaling from 
these considerations (i.e. $L_{bb}=L_{ff}$) is the same scaling as that
for the maximum mass of a mirror star\cite{teplitz} which comes from
finding the mass of a
star such that radiatuion pressure dominates matter pressure thereby
giving rise to instability. It is also the same scaling that one gets from
computing the minimum mass of a mirror star for which it will burn as
well as for the Chandrasekhar mass.
It is intriguing that, even though the physical inputs to these four
calculations are quite different, the resulting scaling law for the
stellar mass at issue (maximum, minimum and Chandrasekhar) is still the
same, i.e. $\zeta^{-2}$.

\end{document}